\documentclass[a4paper,12pt]{article}

\usepackage[left=2cm,top=2cm,right=2cm,bottom=2cm,nohead]{geometry}
\usepackage[format=plain,margin=0.5cm,textformat=simple,justification=justified,small,parskip=0cm]{caption}
\usepackage[pdftex]{graphicx}
\usepackage[pdftex,unicode,pdfstartview=FitH,colorlinks=true,pdfborder={0 0 0 0},urlcolor=blue]{hyperref}
\usepackage{amsmath}

\newcommand{\upd}{\mathrm{d}}
\newcommand{\etal}{\textit{et al.}}

\title{Three-state herding model of the financial markets}
\author{A. Kononovicius, V. Gontis}
\date{}

\begin{document}

\maketitle

\begin{abstract}
We propose a Markov jump process with the three-state herding interaction. We see our approach as an agent-based model for the financial markets. Under certain assumptions this agent-based model can be related to the stochastic description exhibiting sophisticated statistical features. Along with power-law probability density function of the absolute returns we are able to reproduce the fractured power spectral density, which is observed in the high-frequency financial market data. Given example of consistent agent-based and stochastic modeling will provide background for the further developments in the research of complex social systems.

\end{abstract}

\section{Introduction}
Following the recent financial crisis a vast number of papers suggesting what could be improved in the financial policy making and Economics itself were published. In these papers agent-based modeling was seen as one of the key tools, which could improve the understanding of the financial markets as well as lead to the potential applications \cite{Bouchaud2008Nature, Farmer2009Nature, Lux2009NaturePhys}. The idea is not so new, see a paper by Axelrod published back in the 1997, ref. \cite{Axelrod1997Comp}, but it has been actively developed only in the last few years and is seen as one of the potential future prospects \cite{Roehner2010SciCul}.

Currently there are many differing agent-based approaches in the modeling of the financial markets \cite{Cristelli2010Fermi,Chakraborti2011RQUF2}. Reference \cite{Cristelli2010Fermi} suggests that the ideal financial market model should be both realistic, namely include realistic individual trader behavior (with the best example being \cite{Lux1999Nature}), and tractable, namely have an analytical solution (ex. see an analytical solution of the Minority Game by Challet \etal \cite{Challet2000PhysRevLett}). This raises an interesting point - one has to build bridges between microscopic and macroscopic modeling, because doing so might lead to the ideal model. And this actually currently being done. In the recent paper Krause \etal \cite{Krause2012PhysRevE} proposed a macroscopic stochastic model analogous to the Ising model interpretation for the financial markets, introduced in \cite{Bornholdt2001IJMPC}. Another interesting approach was made by Feng \etal \cite{Feng2012PNAS} who have used empirical observations and trader survey data to construct agent-based and stochastic models for the financial markets.

Our approach is based on the integration of two alternatives. One of them is a very simple, yet very relevant and highly applicable \cite{Aoki2007Cambridge,Kononovicius2012IntSys}, Markov jump process based on the Kirman's agent-based herding model, introduced in \cite{Kirman1993QJE}. And the other one is a very general stochastic model \cite{Kaulakys2004PhysRevE,Kaulakys2006PhysA,Ruseckas2010PhysRevE}, which was built for the modeling of return and trading activity in the financial markets \cite{Gontis2008PhysA, Gontis2010PhysA, Gontis2010Intech}.

In the previous articles \cite{Kononovicius2012PhysA,Ruseckas2011EPL} we have used the Kirman's herding transition rates to derive a stochastic model for the absolute returns. In this approach we generalized and extended the work by Alfarano \etal \cite{Alfarano2005CompEco,Alfarano2008JEcoDyC}. By doing so we have established the relations between the Markov jump process and a very general class of stochastic equations,
\begin{equation}
\upd x = \left( \eta - \frac{\lambda}{2} \right) x^{2 \eta -1} \upd t_s + x^\eta \upd W_s , \label{eq:xsde}
\end{equation}
generating power-law statistics. Namely the time series obtained by solving eq. \eqref{eq:xsde} posses power-law stationary probability density and power spectral density as follows:
\begin{equation}
p(x) \sim x^{-\lambda}, \quad S(f) \sim 1/f^\beta,\,\, \beta = 1 + \frac{\lambda-3}{2 (\eta -1)}. \label{eq:xsdefits}
\end{equation}
Stochastic differential equation \eqref{eq:xsde} was previously derived from the point processes and its ability to reproduce power-law statistics was grounded in \cite{Kaulakys2004PhysRevE,Kaulakys2006PhysA,Ruseckas2010PhysRevE}.
Many physical, physiological, and social systems are characterized by complex interactions among different components and power-law correlations in the output of these systems \cite{Kobayashi1982BioMed, Ivanov1998EPL, Ashkenazy2001PhysRevLett, Ashkenazy2002PhysA, Ivanov2004PhysRevE, Podobnik2009PNAS}.
The applications of such stochastic model might include varying complex systems possessing power-law statistical features.

In this paper we considerably extend agent-based herding model introducing the Markov jumps between three-states available to agents (these states may be alternatively seen as agent groups). This approach  lets us reproduce the fractured power spectral density, which is an important statistical feature of the high-frequency empirical financial market data \cite{Gontis2008PhysA, Gontis2010PhysA, Gontis2010Intech, Liu1999PhysRevE}. First we introduce a possible relation between the Markov jump process and power-law stochastic processes, define herding interaction between three agent groups, introduce possible application for the financial markets. Next we give a stochastic treatment for the new approach and finally we discuss results in the context of microscopic and macroscopic modeling.

\section{The Markov jump process in the background of the power-law stochastic processes}
We choose the Markov jump process as a basic stochastic process, which enables the reproduction of the agent dynamics on the microscopic scale. This approach is already in use in restructuring Macroeconomics \cite{Aoki2007Cambridge} and building microscopic models for the financial markets \cite{Kononovicius2012IntSys}. The method works as a more detailed reasoning for the microscopic behavior is irrelevant in determining the macroscopic description.

Let us start from the model with $N$ agents facing binary choice (0 or 1). The state variable $X$ can be considered as a number of agents making choice $1$. In this case the state space is then $\{0,1,2,...,N\}$. One can interpret this model as a random walk, or a birth-death process, because $X$ changes at most by $\pm 1$ in sufficiently small time interval $\Delta t$. The transition probabilities from state $X$ to the states $X\pm1$ may be specified by
\begin{eqnarray}
& p (X+1) = (N-X) \mu_1(X,N) \Delta t ,\label{eq:pgen1}\\
& p (X-1) = X \mu_2(X,N) \Delta t  ,\label{eq:pgen2}
\end{eqnarray}
where $\mu_1(X,N)$ and $\mu_2(X,N)$ are positive transition rate functions. The transition probabilities \eqref{eq:pgen1} and \eqref{eq:pgen2} appear general enough to provide wide opportunities. The most simple one, when the functions $\mu_{1,2}$ are constant, represents well-known birth-death or entry-exit process.

One can obtain the power-law statistics starting from the ref. \cite{Kirman1993QJE}, where Kirman has noticed that a very similar patterns are observed in a relatively different systems. Apparently statistically similar herding behavior is observed in a very different fields - economics, see, e. g., a paper by Becker \cite{Becker1991JPolitEco}, and entomology, the credits goes to Deneubourg and Pastels (see \cite{Detrain2006PhysLifeRev} for the most recent work). The entomological observations concluded that even if the ant colony has two identical food sources available ants still prefer to use only one of them at a given time. The other food source is not completely neglected as the ants after some time switch to it. The economists observe similar behavior - people tend to choose more popular product, than less popular, despite both being of a similar quality.

Taking the discussed empirical observations into account Kirman has proposed to model the herding behavior as a Markovian chain with the following one step transition rates:
\begin{equation}
\mu_1(X,N) = \sigma_1 + h X , \quad \mu_2(X,N) = \sigma_2 + h (N-X) ,\label{eq:Krates}
\end{equation}
here $h$ parameter defines herding behavior, as a property of the agents describing the strength of imitation tendencies. While $\sigma_i$ parameters describe an asymmetric individual transitions of the agents made independently of the other agents' behavior. We will demonstrate that this simple model of herding interaction of agents can be considered as the background of pretty complex power-law behavior of financial variables.

In order to reproduce the sophisticated power-law behavior of the absolute returns in the financial markets we generalized the herding model by assuming that the meeting rates of the agents are not constant, but depend on the system state, $\frac{1}{\tau(X,N)}$, \cite{Kononovicius2012PhysA}. This strengthens a feedback of the macroscopic state on the agent transition rates as follows:
\begin{eqnarray}
& \mu_1(X,N) = \sigma_1 + \frac{h X}{\tau(X,N)} ,\label{eq:pPlus}\\
& \mu_2(X,N) = \frac{\sigma_2 + h (N-X)}{\tau(X,N)} ,\label{eq:pMinus}
\end{eqnarray}
Note that $\sigma_1$ is not divided by $\tau(X,N)$. The reason behind this is a very simple one - we have previously assumed that a modeled market contains $N-X$ rational and long term fundamentalist traders, whose individual behavior should not depend on the temporary fads and moods. It is possible to use other assumptions as the model is rather flexible.

\begin{figure}[!t]
\centering
\includegraphics[width=0.4\textwidth]{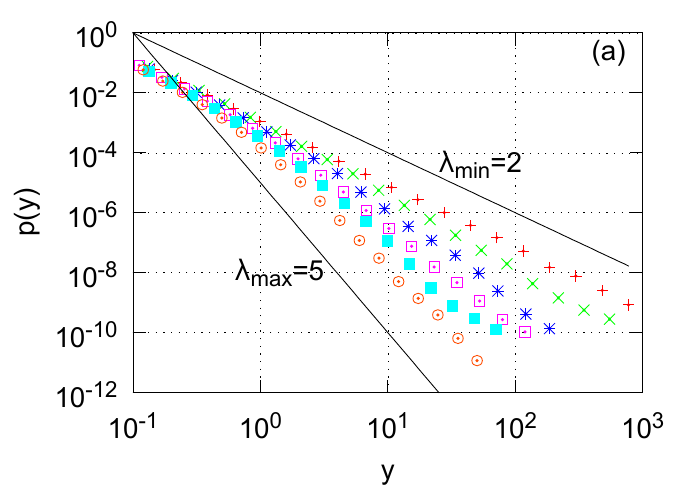}
\hspace{0.05\textwidth}
\includegraphics[width=0.4\textwidth]{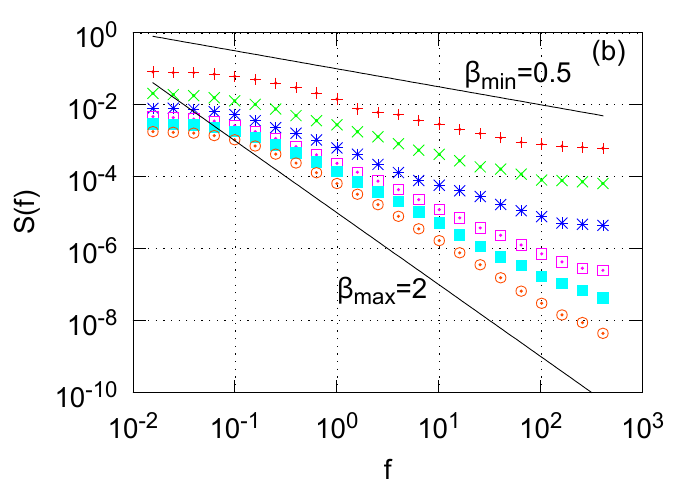}
\caption{Wide spectra of obtainable probability (a) and spectral (b) density functions of stochastic variable $y$, dynamics of which are defined by eq. \eqref{eq:ysdefull}. Black curves are the limiting, minimum and maximum exponent, power-law functions with: (a) $\lambda_{min} = 2$ and $\lambda_{max} = 5$, (b) $\beta_{min} = 0.5$ and $\beta_{max} =2$. Model parameters: $\alpha=1$, $\varepsilon_1 = 0.1$, $\varepsilon_2 = 0.1$ (red plus), $0.5$ (green cross), $1$ (blue stars), $1.5$ (magenta open squares), $2$ (cyan filled squares) and $3$ (orange open circles).}
\label{fig:ysde}
\end{figure}

The further stochastic treatments of the model, mainly relying on the van Kampen birth-death process formalism \cite{VanKampen1992NorthHolland} and the Ito rules for variable substitution \cite{Gardiner1997Springer}, leads to the following stochastic differential equation for new stochastic variable, $y=\frac{X}{N-X}$, introduced as a measure of the absolute return \cite{Kononovicius2012PhysA}:
\begin{equation}
\upd y = \left[ \varepsilon_1 + y \frac{2-\varepsilon_2}{\tau(y)} \right] (1+y) \upd t_s + \sqrt{\frac{2 y}{\tau(y)}} (1+y) \upd W_s , \label{eq:ysdefull}
\end{equation}
here $t_s = h t$. This stochastic differential equation in the limit of large $y$, $y \gg 1$, can be considered to include only the highest powers of $y$. In such case, and by assuming that $\tau(y) = y^{-\alpha}$, one obtains:
\begin{equation}
\upd y = (2-\varepsilon_2) y^{2+\alpha} \upd t_s + \sqrt{2 y^{3+\alpha}} \upd W_s , \label{eq:ysde}
\end{equation}
which is identical to eq. \eqref{eq:xsde}. The direct comparison of eqs. \eqref{eq:xsde} and \eqref{eq:ysde} yields the relation between the models' parameters:
\begin{equation}
\eta = \frac{3+\alpha}{2} , \quad \lambda = \varepsilon_2 + \alpha + 1 .\label{eq:ysdecomp}
\end{equation}

The direct consequence of the comparison is the ability to control the power-law exponents, $\lambda$ and $\beta$, of the $y$ statistical features obtained from the agent-based model, eqs. \eqref{eq:pPlus} and \eqref{eq:pMinus}, and its stochastic treatment, eq. \eqref{eq:ysdefull}. This can be used to reproduce $1/f^\beta$ noise with $0.5 < \beta <2$ (see fig. \ref{fig:ysde}). Yet the most important result is the agent-based reasoning being provided for a very general class of power-law stochastic processes, reproducible  by stochastic differential equations \eqref{eq:xsde}, derived from the point processes \cite{Kaulakys2004PhysRevE,Gontis2004PhysA343,Kaulakys2006PhysA,Ruseckas2010PhysRevE}.

\section{Three-state model with herding interaction}
\begin{figure}[!t]
\centering
\includegraphics[width=0.4\textwidth]{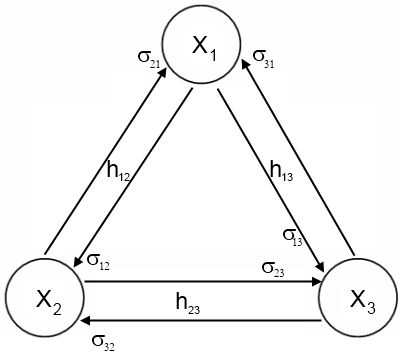}
\caption{Schematic representation of the three-state model. The arrows point in the directions of the possible transitions, note that they can be grouped into three pairs. In the model each of the transition pairs is modeled using the original Kirman's model. The relevant parameters are grouped around the corresponding pairs of arrows.}
\label{fig:schemaModel}
\end{figure}

 In this section we extend the herding model by introducing the three-state agent dynamics. One can easily extend the model by assuming that the original Kirman's transition probabilities describe the transitions between each pair of the agent states in the system. Thus in three-state case with numbers of agents in each group $X_1$, $X_2$, $X_3$ we will have six one step transition probabilities of a general form given by:
\begin{equation}
p (X_i+1, X_j-1, X_k ) = X_j (\sigma_{ji} + h_{ji} X_i) \Delta t .
\end{equation}
The above holds for non-equal $i$, $j$ and $k$ each taking a value from the set $\{1, 2, 3\}$. Note that herding behavior is assumed to be symmetrical, thus we have $h_{ij} = h_{ji}$. The schematic representation of the extended model is given in fig. \ref{fig:schemaModel}.

Note that due to conservation of total number of agents $N$, which is given by $N=X_1+X_2+X_3$, one can fully describe the three-state system by the two dimensional state vector, $\vec X = \{X_1, X_2\}$. We assume that $N$ is large enough to secure the continuity of $x_i = X_i/N$ and introduce the definition of the transition probability densities, $\pi^{i,j}(x_1, x_2)$ :
\begin{equation}
p(X_1 + i, X_2 + j, X_3 + k) = N^2 \pi^{i,j}(x_1, x_2) \Delta t .
\end{equation}
Here the indexes $i$, $j$ and $k$ stand for the corresponding change of the number of agents in the corresponding populations. The indexes $i$, $j$ and $k$ must take different values from the set $\{-1,0,1\}$. Consequently, $i+j+k=0$ and vector $\{i,j,k\}$ can be fully described by its two components $\{i,j\}$.

The transition probabilities imply the Master equation for the probability $\omega(x_1, x_2 ,t)$ to find the system in the state $\{x_1, x_2\}$ at given time $t$:
\begin{equation}
\partial_t \omega = \sum_{i \neq j} \left( {\bf E}^{i,j}-1\right) \pi^{-i,-j} \omega , \label{eq:master}
\end{equation}
here $i$ and $j$ are two non-equal indexes, which take values from the set $\{-1,0,1\}$. In the above ${\bf E}^{i,j}$ is the two variables one step operator, which is a convenient generalization of the one variable one step operator used by van Kampen in \cite{VanKampen1992NorthHolland}. The form of the two variables one step operator is given by:
\begin{equation}
{\bf E}^{i,j} [f(x,y)] = f(x+ i \Delta x, y + j \Delta y) ,
\end{equation}
here $i\neq j \in \{-1,0,1\}$, $\Delta x$ and $\Delta y$ are the smallest possible increments of $x$ and $y$ respectively. Note that the operator acts on all functions on its right side.

We expand this one step operator, using the Tailor series in the limit of small increments, $\Delta x$ and $\Delta y$, up to the second order terms. By recalling that in our case $\Delta x = \Delta y = N^{-1}$ (it follows from the definition of the almost continuous $x_i$) we obtain a Fokker-Plank equation,
\begin{equation}
\partial_t \omega = -\sum_{i} \partial_{x_i} \left[ D_i^1 \omega \right] + \sum_{i , j} \partial_{x_i} \left\{ \partial_{x_j} \left[ D_{ij}^2 \omega \right] \right\}, \label{eq:fokplank}
\end{equation}
where $i$ and $j$ belong to the set $\{1,2\}$, and
\begin{eqnarray}
& D_1^1 = \sigma_{21} x_2 + \sigma_{31} (1-x_2-x_1) - (\sigma_{12} + \sigma_{13}) x_1 , \nonumber \\
& D_2^1 = \sigma_{12} x_1 + \sigma_{32} (1-x_2-x_1) - (\sigma_{21} + \sigma_{23}) x_2 , \nonumber  \\
& D_{11}^2 \approx h_{12} x_1 x_2 + h_{13} x_1 (1-x_2-x_1)  , \label{eq:fpterms} \\
& D_{22}^2 \approx h_{12} x_1 x_2 + h_{23} x_2 (1-x_2-x_1) , \nonumber  \\
& D_{12}^2 = D_{21}^2 \approx - h_{12} x_1 x_2 . \nonumber
\end{eqnarray}
The obtained Fokker-Plank equation appears to be complex, but it can be simplified after some additional assumptions. The financial market interpretation based on the three-state model also enables us to make smooth transition to the system of stochastic differential, Langevin, equations.

\section{Financial market model with the three agent groups}

Let us start by assuming that the three-states available to the agents in the aforementioned setup correspond to the three types of traders: fundamentalists, chartists optimists and chartists pessimists. In the current agent-based modeling it is one of the most common choices \cite{Cristelli2010Fermi,Feng2012PNAS}.

Fundamentalists are the traders who have fundamental understanding of the true value of the traded stock. This understanding is quantified as the stocks fundamental price, $P_f(t)$. For a mathematical convenience and without loosing generality one can assume that the fundamental price does not vary with time. In other words, we are interested in the price fluctuations according to its fundamental value. Having this knowledge available to them the fundamentalists make rational long term expectations. Thus their excess demand, $ED_f(t)$, is given by \cite{Alfarano2005CompEco}:
\begin{equation}
ED_f(t) = N_f(t) \ln \frac{P_f}{P(t)} ,
\end{equation}
where $N_f(t)$ is a number of the fundamentalists inside the market and $P(t)$ is a current market price. The mathematical expression for the excess demand of the fundamentalist traders can be read as follows: if $P_f < P(t)$, the fundamentalist will sell the stock as he expects a decrease of price, while in the opposite case, $P_f > P(t)$, he will buy stock expecting price growth. This behavior is based on the assumption that $P(t)$ should converge towards $P_f$ given enough time.

The other two types, pessimistic and optimistic chartists, are short term traders, who estimate the future price based on its recent movements. Namely these traders rely on the technical trading strategies and short term opinion fluctuations. As there is a wide selection of such strategies and opinions, one can simply generalize by assuming that some strategies and opinions at a given moment are optimistic, i.e. suggesting to buy, while the others are pessimistic, i.e. suggesting to sell. In such case the excess demand of the chartist traders, $ED_c(t)$, is given by:
\begin{equation}
ED_c(t) = r_0 [N_o(t) - N_p(t)] ,
\end{equation}
where $r_0$ is a relative impact factor of the chartist trader, $N_o$ and $N_p$ are the total numbers of optimists and pessimists respectively. In the previous approaches \cite{Kononovicius2012PhysA, Alfarano2005CompEco} chartist traders were considered to be a single group, opinion switching in which was assumed to be purely random, thus over-simplifying the endogenous mood dynamics.

Price and later returns can be introduce into the model by applying the Walrasian scenario. As a fair price is assumed to reflect the current supply and demand, the Walrasian scenario in its contemporary form may be expressed as:
\begin{equation}
\frac{1}{\beta N} \frac{\upd p(t)}{\upd t} = - n_f(t) p(t) + r_0 [n_o(t) - n_p(t)] ,
\end{equation}
here $\beta$ is a speed of the price adjustment, $N$ a total number of traders in the market, $p(t) = \ln \frac{P(t)}{P_f}$ and $n_i(t) = \frac{N_i(t)}{N}$. By assuming that the number of traders in the market is large, $N \rightarrow \infty$, one obtains:
\begin{equation}
p(t) = r_0 \frac{n_o(t) - n_p(t)}{n_f(t)} .
\end{equation}
Consequently the expression of the return in the selected time window $T$ is given by
\begin{equation}
r(t) = r_0 \left[ \frac{n_o(t) - n_p(t)}{n_f(t)} - \frac{n_o(t-T) - n_p(t-T)}{n_f(t-T)} \right] . \label{eq:retfull}
\end{equation}
In the previous approaches \cite{Kononovicius2012PhysA, Alfarano2005CompEco} the expression for the returns, eq. \eqref{eq:retfull}, was simplified by assuming that the chartist traders change their opinion significantly faster than the fundamentalist traders. Further in this work we will use this assumption as well as regarding eq. \eqref{eq:retfull} as a definition for the return.

Note that in the above discussion we introduced some assumptions about the three agent states, which ought to be considered in the financial market scenario. We can further develop these ideas and simplify the Fokker-Planck equation obtained in the previous section for the general case of the three-state model. First of all let us link the states with actual agent types:
\begin{equation}
x_1 = n_f , \quad x_2 = n_p , \quad x_3 = n_o .
\end{equation}
Next let us point out the lack of qualitative difference between optimism and pessimism:
\begin{equation}
\sigma_{23} = \sigma_{32} = \sigma_{cc} , \quad \sigma_{12} = \sigma_{13} = \sigma_{fc}/2 , \sigma_{21} = \sigma_{31} = \sigma_{cf}, \quad h_{12} = h_{13} = h_1 .
\end{equation}
Finally let us use the assumption that chartists change opinion significantly faster than fundamentalists:
\begin{equation}
h_{23} = H h_1 , \quad H \gg 1, \quad \sigma_{cc} \gg \sigma_{cf}, \quad \sigma_{cc} \gg \sigma_{fc} ,
\end{equation}
where $H$ is a speed ratio.

Under the financial market scenario the terms of the Fokker-Plank equation derived in previous section, eq. \eqref{eq:fpterms}, can now be re-expressed as:
\begin{eqnarray}
& D_f^1 = \sigma_{cf} (1-n_f) - \sigma_{fc} n_f , \nonumber \\
& D_p^1 \approx \sigma_{cc} (1-n_f- 2 n_p) , \nonumber \\
& D_{ff}^2 \approx h_{1} (1- n_f) n_f  , \label{eq:fptermsFin}\\
& D_{pp}^2 \approx H h_1 n_p (1-n_f-n_p) + h_{1} n_f n_p , \nonumber \\
& D_{fp}^2 = D_{pf}^2 \approx - h_{1} n_f n_p . \nonumber
\end{eqnarray}

\section{The system of stochastic differential equations for the three-state model}

It is useful to derive the stochastic differential equations analogous to the Fokker-Plank equation \eqref{eq:fokplank}. The general form of a stochastic differential equation in the case of two variables can be written as
\begin{equation}
\upd | n \rangle = | R \rangle \upd t + [ {\bf S} \cdot \upd | W \rangle ] ,
\end{equation}
with the state vector $ | n \rangle $, vector of the relaxation functions $ | R \rangle $, matrix of the diffusion functions $\mathbf{S}$, and the vector of Brownian motion $ | W \rangle $.

The elements of the matrix of the diffusion functions, $\bf S$, are related to the second order terms of the Fokker-Plank equation as \cite{Risken1996Springer}
\begin{equation}
D^2_{ij} = \frac{1}{2} \sum_{\forall k} S_{ik} S_{jk} , \quad \forall i, j .\label{eq:D2ij}
\end{equation}
After substituting the $D^2_{ij}$ for the expressions from eqs. \eqref{eq:fptermsFin} we obtain a system of a three linearly independent equations for the four elements of $\bf S$. It is convenient to additionally assume that $S_{fp} = S_{pf}$. Then by solving eqs. \eqref{eq:D2ij} we obtain all of the diffusion functions. Consequently we obtain the system of the stochastic differential equations:
\begin{eqnarray}
& \upd n_f = \left[ (1-n_f) \sigma_{cf} - n_f \sigma_{fc} \right] \upd t + \sqrt{2 h_1 n_f (1-n_f)} \upd W_1 , \\
& \upd n_p = (1-n_f - 2 n_p) \sigma_{cc} \upd t + \sqrt{2 H h_1 n_p (1-n_f-n_p)} \upd W_2 .
\label{eq:SDEfp}
\end{eqnarray}
Interestingly enough similar equations are obtained for the evolutionary three strategy games in the series of papers by Traulsen \etal \cite{Traulsen2005PhysRevLett, Traulsen2006PhysRevE, Traulsen2012PhysRevE}.

The above stochastic differential equations are inter-dependent, while it would be more convenient to have a system of independent stochastic differential equations. Taking into account the previous approaches \cite{Kononovicius2012PhysA, Alfarano2005CompEco} one can expect that the introduction of the mood, $\xi(t) = \frac{n_o(t)-n_p(t)}{n_o(t)+n_p(t)}$, as a new variable instead of $n_p$ would solve this problem. This can be done either via Ito rules for variable substitution in Langevin equation \cite{VanKampen1992NorthHolland} or via variable substitution in the Fokker-Plank equation \cite{Risken1996Springer}.

At this point let us scale the time, $t_s = h_1 t$, and appropriately redefine the model parameters: $\varepsilon_{cf} = \sigma_{cf} / h_1$, $\varepsilon_{fc} = \sigma_{fc} / h_1$, $\varepsilon_{cc} = \sigma_{cc} / (H h_1)$. Let us also recall our generalization of the herding model, eqs. \eqref{eq:pPlus} and \eqref{eq:pMinus}, by introducing the additional variability of the event rate, $\tau(\dots)$. The same assumptions can be used to introduce the variability into the three-state model. In such a case one can get:
\begin{eqnarray}
& \upd n_f = \left[ \frac{(1-n_f) \varepsilon_{cf}}{\tau(n_f,\xi)} - n_f \varepsilon_{fc} \right] \upd t_s + \sqrt{\frac{2 n_f (1-n_f)}{\tau(n_f,\xi)}} \upd W_{s,1} , \label{eq:nftau}\\
& \upd \xi = - \frac{2 H \varepsilon_{cc} \xi}{\tau(n_f,\xi)} \upd t + \sqrt{\frac{2 H (1-\xi^2)}{\tau(n_f,\xi)}} \upd W_{s,2} , \label{eq:xitau}\\
& \tau(n_f,\xi) = \left[ 1 + \left| \frac{1-n_f}{n_f} \xi \right|^\alpha \right]^{-1} . \label{eq:taunfxi}
\end{eqnarray}
Note that in previous approach \cite{Kononovicius2012PhysA} we have defined $\tau(\dots)$ as an inverse function of $y$, namely $\tau(n_f) = y^{-\alpha} = \left[ \frac{n_f}{1-n_f} \right]^\alpha$. This form was selected for the sake of simplicity and also considering the limit of large absolute returns, $y \gg 1$. In the three agent group approach we choose to move further away from this simplification and introduce an inverse dependence on the log-price, $p =\frac{1-n_f}{\xi}$, thus arriving at the eq. \eqref{eq:taunfxi}. Similar approach was used in \cite{Lux1999Nature}, where the utility functions of the agents, and consequently their opinion switching, are dependent on the log-price.

\begin{figure}[!t]
\centering
\includegraphics[width=0.39\textwidth]{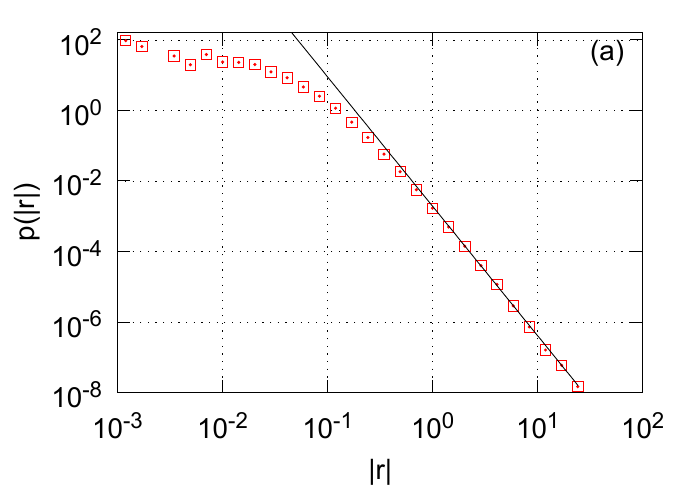}
\hspace{0.05\textwidth}
\includegraphics[width=0.39\textwidth]{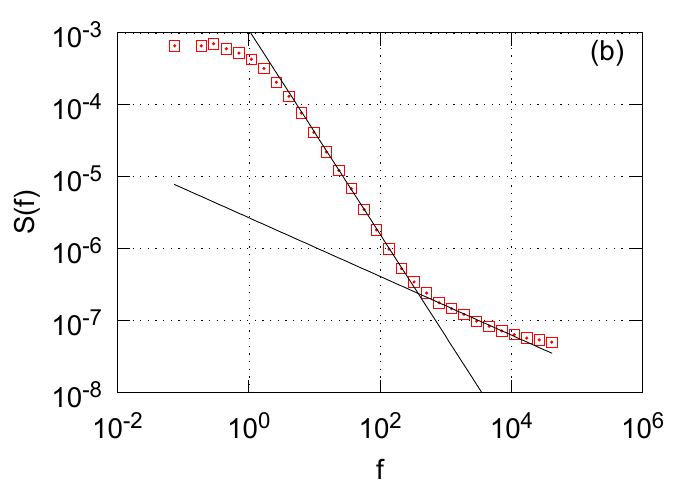}
\caption{Probability density function (a) and power spectral density (b) of absolute return numerically calculated from the three-state agent-based model. The red squares represent the numerical results obtained by solving eqs. \eqref{eq:nftau} and \eqref{eq:xitau}. Model parameters were set as follows: $\varepsilon_{cf} = \varepsilon_{fc} = \varepsilon_{cc}=3$, $H=100$, $r_0=1$, $\alpha=2$. The black curves provide power-law fits: (a) $\lambda = 3.67$, (b) $\beta_1=1.42$ and $\beta_2=0.41$.}
\label{fig:fractured}
\end{figure}

In fig. \ref{fig:fractured} we show that this model possesses a fractured spectral density similar to the one obtained in the sophisticated stochastic models considered in \cite{Gontis2010PhysA, Gontis2010Intech}. The probability density function is a $q$-Gaussian like and has a power-law tail with the close-to-empirical exponent. Though the spectral density is fitted by the power-law functions with larger exponents than the empirical ones. The same behavior was also observed in the stochastic model for the absolute returns. In case of  \cite{Gontis2010PhysA, Gontis2010Intech} the model's spectral density was reconciled with the empirical data by applying the additional $q$-Gaussian noise driven by the resulting time series. In a sense of the proposed stochastic model derived as endogenous fluctuations of three agent groups we can expect that the system's response to the exogenous fluctuations of information flow is defined by endogenous macroscopic state.

\section{Conclusions}

In this work we derived the system of stochastic equations \eqref{eq:SDEfp} modulating dynamics of three agent groups with herding interaction. Proposed approach can be valuable in the modeling of the complex social systems with similar composition of the agents. We demonstrate how sophisticated statistical features of the absolute returns in financial markets can be reproduced by extending the herding interaction of the agents and introducing the third agent state. Fortunately, the model retains its macroscopic treatment by the stochastic differential equations \eqref{eq:nftau} and \eqref{eq:xitau}. Faster transitions between optimism and pessimism introduce two time scales of the model reflected in the fractured power spectral density fig. \ref{fig:fractured} (b). This is in qualitative agreement with the high-frequency empirical data, which exhibits a similar behavior with lower exponents of the spectral density of the absolute returns \cite{Liu1999PhysRevE,Gontis2010Intech}. Agent-based model in the present form considers only the endogenous fluctuations, while the exogenous fluctuations are related to the information flow and would be considered as additional noise. Earlier we proposed a double stochastic model, which demonstrates the influence of additional noise reducing exponents of power spectrum \cite{Gontis2010PhysA,Gontis2010Intech}.

\section*{Acknowledgments}

We would like to express our gratitude to dr. Julius Ruseckas who provided useful suggestions and advices.

\bibliographystyle{IEEEtr}
\bibliography{physrisk}

\begin{thebibliography}{10}

\bibitem{Bouchaud2008Nature}
J.~P. Bouchaud, ``Economics need a scientific revolution,'' {\em Nature},
  vol.~455, p.~1181, 2008.

\bibitem{Farmer2009Nature}
J.~D. Farmer and D.~Foley, ``The economy needs agent-based modelling,'' {\em
  Nature}, vol.~460, pp.~685--685, 2009.

\bibitem{Lux2009NaturePhys}
T.~Lux and F.~Westerhoff, ``Economic crysis,'' {\em Nature Physics}, vol.~5,
  pp.~2--3, 2009.

\bibitem{Axelrod1997Comp}
R.~Axelrod, ``Advancing the art of simulation in the social sciences,'' {\em
  Complexity}, vol.~3, no.~2, pp.~16--32, 1997.

\bibitem{Roehner2010SciCul}
B.~M. Roehner, ``Fifteen years of econophysics: worries, hopes and prospects,''
  {\em Science and culture}, vol.~76, pp.~305--314, 2010.

\bibitem{Cristelli2010Fermi}
M.~Cristelli, L.~Pietronero, and A.~Zaccaria, ``Critical overview of
  agent-based models for economics,'' in {\em Proceedings of the School of
  Physics {"}E. Fermi{"}, Course CLXXVI} (F.~Mallnace and H.~E. Stanley, eds.),
  (Bologna-Amsterdam), pp.~235 -- 282, SIF-IOS, 2012.

\bibitem{Chakraborti2011RQUF2}
A.~Chakraborti, I.~M. Toke, M.~Patriarca, and F.~Abergel, ``Econophysics
  review: Ii. agent–based models,'' {\em Quantitative Finance}, vol.~7,
  pp.~1013--1041, 2011.

\bibitem{Lux1999Nature}
T.~Lux and M.~Marchesi, ``Scaling and criticality in a stochastic multi-agent
  model of a financial market,'' {\em Nature}, vol.~397, pp.~498--500, 1999.

\bibitem{Challet2000PhysRevLett}
D.~Challet, M.~Marsili, and R.~Zecchina, ``Statistical mechanics of systems
  with heterogeneous agents: Minority games,'' {\em Physical Review Letters},
  vol.~84, no.~8, pp.~1824--1827, 2000.

\bibitem{Krause2012PhysRevE}
S.~M. Krause, P.~Bottcher, and S.~Bornholdt, ``Mean-field-like behavior of the
  generalized voter-model-class kinetic ising model,'' {\em Physical Review E},
  vol.~85, p.~031126, 2012.

\bibitem{Bornholdt2001IJMPC}
S.~Bornholdt, ``Expectation bubbles in a spin model of markets: Intermittency
  from frustration across scales,'' {\em International Journal of Modern
  Physics C}, vol.~12, no.~5, pp.~667--674, 2001.

\bibitem{Feng2012PNAS}
L.~Feng, B.~Li, B.~Podobnik, T.~Preis, and H.~E. Stanley, ``Linking agent-based
  models and stochastic models of financial markets,'' {\em Proceedings of the
  National Academy of Sciences of the United States of America},
  pp.~1205013109v1--6, 2012.

\bibitem{Aoki2007Cambridge}
M.~Aoki and H.~Yoshikawa, {\em Reconstructing Macroeconomics A Perspektive from
  Statistical Physics and Combinatorial Stochastic Processes}.
\newblock Cambridge University Press, 2007.

\bibitem{Kononovicius2012IntSys}
A.~Kononovicius, V.~Gontis, and V.~Daniunas, ``Agent-based versus macroscopic
  modeling of competition and business processes in economics and finance,''
  {\em International Journal On Advances in Intelligent Systems}, vol.~5,
  no.~1-2, pp.~111--126, 2012.

\bibitem{Kirman1993QJE}
A.~P. Kirman, ``Ants, rationality and recruitment,'' {\em Quarterly Journal of
  Economics}, vol.~108, pp.~137--156, 1993.

\bibitem{Kaulakys2004PhysRevE}
B.~Kaulakys and J.~Ruseckas, ``Stochastic nonlinear differential equation
  generating 1/f noise,'' {\em Physical Review E}, vol.~70, no.~2, p.~020101,
  2004.

\bibitem{Kaulakys2006PhysA}
B.~Kaulakys, J.~Ruseckas, V.~Gontis, and M.~Alaburda, ``Nonlinear stochastic
  models of 1/f noise and power-law distributions,'' {\em Physica A}, vol.~365,
  pp.~217--221, 2006.

\bibitem{Ruseckas2010PhysRevE}
J.~Ruseckas and B.~Kaulakys, ``1/f noise from nonlinear stochastic differential
  equations,'' {\em Physical Review E}, vol.~81, p.~031105, 2010.

\bibitem{Gontis2008PhysA}
V.~Gontis, B.~Kaulakys, and J.~Ruseckas, ``Trading activity as driven poisson
  process: comparison with empirical data,'' {\em Physica A}, vol.~387,
  pp.~3891--3896, 2008.

\bibitem{Gontis2010PhysA}
V.~Gontis, J.~Ruseckas, and A.~Kononovicius, ``A long-range memory stochastic
  model of the return in financial markets,'' {\em Physica A}, vol.~389,
  pp.~100--106, 2010.

\bibitem{Gontis2010Intech}
V.~Gontis, J.~Ruseckas, and A.~Kononovicius, ``A non-linear stochastic model of
  return in financial markets,'' in {\em Stochastic Control} (C.~Myers, ed.),
  Intech, 2010.

\bibitem{Kononovicius2012PhysA}
A.~Kononovicius and V.~Gontis, ``Agent based reasoning for the non-linear
  stochastic models of long-range memory,'' {\em Physica A}, vol.~391, no.~4,
  pp.~1309--1314, 2012.

\bibitem{Ruseckas2011EPL}
J.~Ruseckas, B.~Kaulakys, and V.~Gontis, ``Herding model and 1/f noise,'' {\em
  EPL}, vol.~96, p.~60007, 2011.

\bibitem{Alfarano2005CompEco}
S.~Alfarano, T.~Lux, and F.~Wagner, ``Estimation of agent-based models: The
  case of an asymmetric herding model,'' {\em Computational Economics},
  vol.~26, no.~1, pp.~19--49, 2005.

\bibitem{Alfarano2008JEcoDyC}
S.~Alfarano, T.~Lux, and F.~Wagner, ``Time variation of higher moments in a
  financial market with heterogeneous agents: An analytical approach,'' {\em
  Journal of Economic Dynamics and Control}, vol.~32, pp.~101--136, 2008.

\bibitem{Kobayashi1982BioMed}
M.~Kobayashi and T.~Musha, ``1/f fluctuation of heartbeat period,'' {\em IEEE
  Transactions on Biomedical Engineering}, vol.~29, pp.~456--457, 1982.

\bibitem{Ivanov1998EPL}
P.~C. Ivanov, L.~A.~N. Amaral, A.~L. Goldberger, and H.~E. Stanley,
  ``Stochastic feedback and the regulation of biological rhythms,'' {\em EPL},
  vol.~43, no.~4, pp.~363--368, 1998.

\bibitem{Ashkenazy2001PhysRevLett}
Y.~Ashkenazy, P.~C. Ivanov, S.~Havlin, C.-K. Peng, A.~L. Goldberger, and H.~E.
  Stanley, ``Magnitude and sign correlations in heartbeat fluctuations,'' {\em
  Physical Review Letters}, vol.~86, pp.~1900--1903, 2001.

\bibitem{Ashkenazy2002PhysA}
Y.~Ashkenazy, J.~M. Hausdorff, P.~C. Ivanov, and H.~E. Stanley, ``A stochastic
  model of human gait dynamics,'' {\em Physica A}, vol.~316, pp.~662--670,
  2002.

\bibitem{Ivanov2004PhysRevE}
P.~C. Ivanov, A.~Yuen, B.~Podobnik, and Y.~Lee, ``Common scaling patterns in
  intertrade times of u. s. stocks,'' {\em Physical Review E}, vol.~69, no.~5,
  p.~056107, 2004.

\bibitem{Podobnik2009PNAS}
B.~Podobnik, D.~Horvati, A.~M. Petersen, and H.~E. Stanley,
  ``Cross-correlations between volume change and price change,'' {\em
  Proceedings of the National Academy of Sciences of the United States of
  America}, vol.~106, pp.~22079--22084, 2009.

\bibitem{Liu1999PhysRevE}
Y.~Liu, P.~Gopikrishnan, P.~Cizeau, M.~Meyer, C.-K. Peng, and H.~E. Stanley,
  ``The statistical properties of the volatility of price fluctuations,'' {\em
  Physical Review E}, vol.~60, pp.~1390--1400, 1999.

\bibitem{Becker1991JPolitEco}
G.~S. Becker, ``A note on restaurant pricing and other examples of social
  influence on price,'' {\em Journal of Political Economy}, vol.~99,
  pp.~1109--1116, 1991.

\bibitem{Detrain2006PhysLifeRev}
C.~Detrain and J.~L. Deneubourg, ``Self-organized structures in a
  superorganism: do ants behave like molecules?,'' {\em Physics of Life
  Reviews}, vol.~3, pp.~162--187, 2006.

\bibitem{VanKampen1992NorthHolland}
N.~G. van Kampen, {\em Stochastic process in Physics and Chemistry}.
\newblock Amsterdam: North Holland, 1992.

\bibitem{Gardiner1997Springer}
C.~W. Gardiner, {\em Handbook of stochastic methods}.
\newblock Berlin: Springer, 1997.

\bibitem{Gontis2004PhysA343}
V.~Gontis and B.~Kaulakys, ``Multiplicative point process as a model of trading
  activity,'' {\em Physica A}, vol.~343, pp.~505--514, 2004.

\bibitem{Risken1996Springer}
H.~Risken, {\em The Fokker-Planck Equation: Methods of Solutions and
  Applications}.
\newblock Springer, 3~ed., 1996.

\bibitem{Traulsen2005PhysRevLett}
A.~Traulsen, J.~C. Claussen, and C.~Hauert, ``Coevolutionary dynamics: From
  finite to infinite populations,'' {\em Physical Review Letters}, vol.~95,
  p.~238701, 2005.

\bibitem{Traulsen2006PhysRevE}
A.~Traulsen, J.~C. Claussen, and C.~Hauert, ``Coevolutionary dynamics in large,
  but finite populations,'' {\em Physical Review E}, vol.~74, p.~011901, 2006.

\bibitem{Traulsen2012PhysRevE}
A.~Traulsen, J.~C. Claussen, and C.~Hauert, ``Stochastic differential equations
  for evolutionary dynamics with demographic noise and mutations,'' {\em
  Physical Review E}, vol.~85, p.~041901, 2012.

\end{thebibliography}

\end{document}